\documentclass{ws-mpla}
\usepackage[super]{cite}
\usepackage{graphicx}
\usepackage{url}
\usepackage{amsmath}
\usepackage{graphicx}

\usepackage{hyperref}\hypersetup{colorlinks=true}

\begin{document}

\newcommand{\be}{\begin{equation}}
\newcommand{\ee}{\end{equation}}
\newcommand{\dd}{{\rm d}}
\newcommand{\snumb}{{\cal N}}
\newcommand{\etal}{{\it et al.}}
\newcommand{\obs}{{\rm o}}

\markboth{Thiago S. Pereira, Davincy T. Pabon}
{Extending the $\Lambda$CDM model through shear-free anisotropies}

%%%%%%%%%%%%%%%%%%%%% Publisher's Area please ignore %%%%%%%%%%%%%%
\catchline{}{}{}{}{}
%%%%%%%%%%%%%%%%%%%%%%%%%%%%%%%%%%%%%%%%%%%%%%%%%%%%%%%%%%%%%%%%%%%

\title{EXTENDING THE $\Lambda$CDM MODEL THROUGH SHEAR-FREE ANISOTROPIES}

\author{\footnotesize THIAGO S. PEREIRA and DAVINCY T. PABON
%\footnote{
%Typeset names in 8 pt Times Roman, uppercase. Use the footnote to
%indicate the present or permanent address of the author.}
}

\address{Departamento de F\'isica, Universidade Estadual de Londrina\\ 
Rodovia Celso Garcia Cid, km 380, 86057-970\\
Londrina -- PR,
Brazil
%\footnote{State completely without abbreviations, the
%affiliation and mailing address, including country and e-mail address.
%Typeset in 8 pt Times Italic.}
\\
tspereira@uel.br}

%\author{SECOND AUTHOR}

%\address{Group, Laboratory, Address\\
%City, State ZIP/Zone, Country
%}

\maketitle

\pub{Received (Day Month Year)}{Revised (Day Month Year)}

\begin{abstract}
If the spacetime metric has anisotropic spatial curvature, one can afford to expand the universe 
isotropically, provided that the energy-momentum tensor satisfy a certain constraint. This leads to 
the so-called shear-free metrics, which have the interesting property of violating the cosmological 
principle while still preserving the isotropy of the cosmic microwave background (CMB) radiation. In 
this work we show that shear-free cosmologies correspond to an attractor solution in the space of 
models with anisotropic spatial curvature. Through a rigorous definition of linear perturbation 
theory in these spacetimes, we show that shear-free models represent a viable alternative to 
describe the large-scale evolution of the universe, leading, in particular, to a kinematically 
equivalent Sachs-Wolfe effect. Alternatively, we discuss some specific signatures that shear-free 
models would imprint on the temperature spectrum of CMB.

\end{abstract}

%\ccode{PACS Nos.: include PACS Nos.}
\vspace{0.5cm}

\section{Introduction}	

The standard concordance model of cosmology -- or $\Lambda$CDM model -- is based on three main 
ingredients: i) the validity of general relativity at cosmological scales, ii) the validity of the 
standard model of particle physics at all cosmological epochs and distances and iii) the 
cosmological principle, according to which our universe is, on average, spatially 
homogeneous, isotropic and infinite. While the first two ingredients have been largely tested, 
modified and scrutinized as an attempt to explain diverse phenomena such as dark matter and dark 
energy, attempts to question the validity of the cosmological principle happens at a slower 
pace, mainly because of our inabilities to collect data from regions other than our 
insignificant corner in an otherwise indifferent and colossal universe.

Notwithstanding the fact that the cosmic microwave background (CMB) radiation is isotropic at 
$0.001\%$ level~\cite{Ade:2015xua,Komatsu:2010fb}, and that the 
distribution of matter at scales above 100$h^{-1}$Mpc are compatible with the cosmological 
principle~\cite{Wu:1998ad,Scrimgeour:2012wt,Pandey:2015xea}, one cannot take for granted a 
principle of central importance to the whole scientific endeavor. On the other hand, attempts 
to extend the cosmological principle have to cope with these very data suggesting isotropy and 
homogeneity.

There are two ways of bridging the extensions of the cosmological principle with observational 
data. One is to admit that small deviations of isotropy and homogeneity are hidden under current 
cosmological data, either lurking in the precision of our current experiments or, to take the 
example of CMB, in the form of large-scale statistical anomalies~\cite{Copi:2010na}. The second 
alternative is to formulate symmetry-violating models that respect the data we have at hand.

In this work we explore the second of these alternatives and investigate a class of spatially 
anisotropic models which preserve, at the background level, the observed isotropy of CMB. In 
particular, we start from previous works in which the anisotropy of the universe results from 
the curvature of the spatial sections, and not from the kinematics of 
expansion~\cite{Mimoso:1993ym,Carneiro:2001fz,Carneiro:2002cj,Koivisto:2010dr}. Such feature is 
implemented by metrics admitting shear-free expansion\cite{Mimoso:1993ym}, and in this work we 
focus on two particular cases: Bianchi type III and Kantowski-Sachs metrics.

This work is organized as follows: in section~\ref{background} we describe a simple class of 
anisotropic metrics admitting anisotropic spatial curvature, and show that, for a specific choice 
of the energy-matter content, these metrics lead to an attractor solution in which the universe 
expands isotropically. Assuming that shear-free models go through a period of inflationary 
expansion, we show in section~\ref{perturbation} that the theory of linear perturbations in 
these models is feasible, and leads to very specific signatures. In section~\ref{signatures} we 
explore a few signatures that shear-free models would imprint on the temperature spectrum of 
CMB. We conclude in section~\ref{perspectives} with some perspectives of extensions of this work.

Throughout this work use metric signature $(-,+,+,+)$ and adopt units such that 
$c=1=8\pi G$. Space and spacetime indices are represented by Latin and Greek letters, 
respectively. The lower case letters $(a,b,c)$ represent coordinates on two-dimensional 
manifolds.

\section{Shear-free anisotropy}

Once we are willing to admit our ignorance about the global symmetries of the universe, we find 
that there is much more to anisotropy than just anisotropic 
expansion~\cite{Barrow:1997mj,Barrow:1998ih,Barrow:1997sy}. In fact, there exists anisotropic 
solutions of the Einstein field equations in which not only the expansion of the universe is 
anisotropic, but so is the curvature of spatial 
sections~\cite{gron2007einstein,ellis2012relativistic}. In the standard four-dimensional description 
of the universe, one way\footnote{Evidently, there exist more sophisticated 
three-dimensional geometries, such as the \textsf{Nil}, \textsf{Sol} and 
$SL(2,\mathbb{R})$ geometries, which we will not consider here. For a recent cosmological study 
involving a Bianchi type II solution (which corresponds to the \textsf{Nil} geometry) 
see Ref.~\refcite{Hervik:2011xm}.} of constructing a manifold with anisotropic curvature is by 
multiplying the (flat) one-dimensional real line $\mathbb{R}$ with a curved two-dimensional space 
$\mathbb{M}$. If we restrict, for the sake of simplicity, to maximally symmetric two-dimensional 
spaces, then there are only two possibilities: either $\mathbb{M}$ is a sphere ($\mathbb{S}^2$) or a 
pseudo sphere ($\mathbb{H}^2$). The first case represents the known Kantowski-Sachs (KS) anisotropic 
solution, while the second gives the Bianchi type III (BIII) metric. In comoving cylindrical 
coordinates, these two solutions can be parameterized as follows:
\be\label{b3ks}
\dd s^2 = -\dd t^2 + 
e^{2\alpha}\left[e^{2\sigma}\left(\dd\rho^2+\frac{1}{|\kappa|}\sin^2(\sqrt{|\kappa|}
\rho)\dd\varphi^2\right)+e^{-4\sigma}\dd z^2\right]\,,
\ee
where $\alpha$ is the average scale factor and $\sigma$ measures the spatial shear. The number 
$\kappa$ measures the curvature of the two-dimensional spaces, and can be either $-1$ (BIII), $+1$ 
(KS). Incidentally, we note that $\kappa=0$ also corresponds to the locally-rotationally-symmetric 
Bianchi I solution.

Given that~\eqref{b3ks} is already anisotropic at the level of the spatial curvature, it is 
natural to ask whether these models can evolve with a single scale factor. Indeed, it has been shown 
that for some specific choices of the energy-momentum content, these models admit a shear-free (SF)
expansion~\cite{Mimoso:1993ym,Carneiro:2001fz}, that is, there exist \emph{anisotropic} 
cosmological solutions with $\sigma=0$ in~\eqref{b3ks}. Thus, before developing the observational 
signatures of SF models, it is important to investigate whether these solutions are dynamically 
stable.

\subsection{Background dynamics}\label{background}

Let us consider the scenario of a 
universe with metric~\eqref{b3ks} and composed of a perfect fluid with energy density $\rho_f$ and 
pressure $p_f$, plus some anisotropic source of energy and momentum. To be more specific, let us 
model the latter by a two-form field $B_{\mu\nu}$, for which we know that shear-free solutions 
exist~\cite{Koivisto:2010dr}\footnote{For a recent application of two-form fields in the context 
of anisotropic cosmologies, see Ref.~\refcite{Ito:2015sxj}}. The total energy-momentum tensor of the 
system is thus:
\begin{align}
T_{\mu\nu} & = (\rho+p)u_\mu u_\nu + p g_{\mu\nu} + \pi_{\mu\nu}\nonumber \\
& = \left[(\rho_f+p_f)u_\mu u_\nu +p_f g_{\mu\nu}\right] + \left[-3\gamma 
J_{\mu\alpha\beta}J^{\alpha\beta}_{\phantom{00}\nu}
+\frac{1}{2}\gamma J_{\alpha\beta\gamma}J^{\alpha\beta\lambda} g_{\mu\nu}\right]\,,
\label{tmunutotal}
  \end{align}
where $\gamma$ is a constant and the field strength 
$J_{\mu\nu\lambda}=3!\partial_{[\mu}B_{\nu\lambda]}$ is such that 
${\partial_\mu(\sqrt{-g}J^{\mu\nu\lambda})=0}$\footnote{We are assuming that the two-form 
field does not couple to the perfect fluid.}. In four dimensions, $J_{\mu\nu\lambda}$ has 
only four components, which means that it is dual to a four vector $V^\rho$. Moreover, since the 
$z$-direction has a distinct character in the coordinates adopted in~\eqref{b3ks}, we will define
\be
\label{jota}
J_{\mu\nu\lambda}\equiv\epsilon_{\mu\nu\lambda\rho}V^{\rho}\,,\qquad V^\rho \equiv 
V\delta^\rho_{\;3}
\ee
where $V$ is a function of time. The Einstein field equations resulting 
from~\eqref{b3ks}-\eqref{jota} are:
\begin{align}
H^{2}-\dot{\sigma}^{2} & =\frac{1}{3}\rho-\frac{R^{(3)}}{6}\label{constraint}\\
\dot{H}+3H^{2} & =\frac{1}{2}\left(\rho-p\right)-\frac{R^{(3)}}{3}\\
\ddot{\sigma}+3H\dot{\sigma} & =\pi_{\perp}-\frac{R^{(3)}}{6}\label{balance}\\
\dot{R}^{(3)} & =-2(H+\dot{\sigma})R^{(3)}\label{dynamic-eq}
\end{align}
where $\rho=\rho_f+\rho_B$ and $p=p_f+p_B$ are the total energy density and pressure, respectively, 
${R^{(3)}=2\kappa 
e^{-2\alpha-2\sigma}}$ is the three-dimensional Ricci scalar, and
\be
\rho_B = -3\gamma e^{2\alpha-4\sigma}V^2\,,\quad p_B = -\frac{1}{3}\rho_B \,,\quad
\pi_\perp\equiv\pi^1_{\;1}=2\gamma e^{2\alpha-4\sigma}V^2\,.
\ee
The fluid variables are also constrained by the equations
\begin{align}
\dot{\rho}_f+3H(\rho_f+p_f) & = 0 \,,\\
\dot{\rho}_B + 3H(\rho_B+p_B) & = -6\dot\sigma\pi_\perp\,,
\end{align}
as follows from the Bianchi identities. Note that, from the positiveness of $\rho_B$, one 
requires $\gamma<0$.

In order to analyze the linear stability of the system it is convenient to work with the following
dimensionless variables
\be
\label{omegas}
\Omega_f\equiv\frac{\rho_f}{3H^2}\,,\quad \Omega_B\equiv\frac{\rho_B}{3H^2}\,,\quad 
\Omega_\kappa\equiv\frac{-R^{(3)}}{6H^2}\,,\quad  \Sigma \equiv \frac{\dot\sigma}{H}\,.
\ee
Note that they are not all independent, but must obey the constraint 
${\Omega_f+\Omega_B+\Omega_k+\Sigma^2=1}$. Eliminating $\Omega_f$ in terms of the other 
variables, the dynamical system becomes
\begin{align*}
\frac{d\Omega_B}{d\alpha} & = 
2\Omega_B\left[3\Sigma^2+2\Sigma+\Omega_B+\Omega_\kappa-1+\frac{3}{2}(1+\omega)
(1-\Sigma^2-\Omega_B-\Omega_\kappa)\right], \\
\frac{d\Omega_\kappa}{d\alpha} & = 
2\Omega_\kappa\left[
3\Sigma^2-\Sigma+\Omega_B+\Omega_\kappa-1+\frac{3}{2}(1+\omega)
(1-\Sigma^2-\Omega_B-\Omega_\kappa)\right],\\
\frac{d\Sigma}{d\alpha} & = 
-2\Omega_B+\Omega_\kappa+\Sigma\left[3(\Sigma^2-1)+\Omega_B+\Omega_\kappa+\frac{3}{2}(1+\omega)
(1-\Sigma^2-\Omega_B-\Omega_\kappa)\right],\nonumber
\end{align*}
where we have assumed that the perfect fluid has an equation of state $\omega=\rho_f/p_f$.

\begin{figure}[th]
\centerline{\includegraphics[scale=0.5]{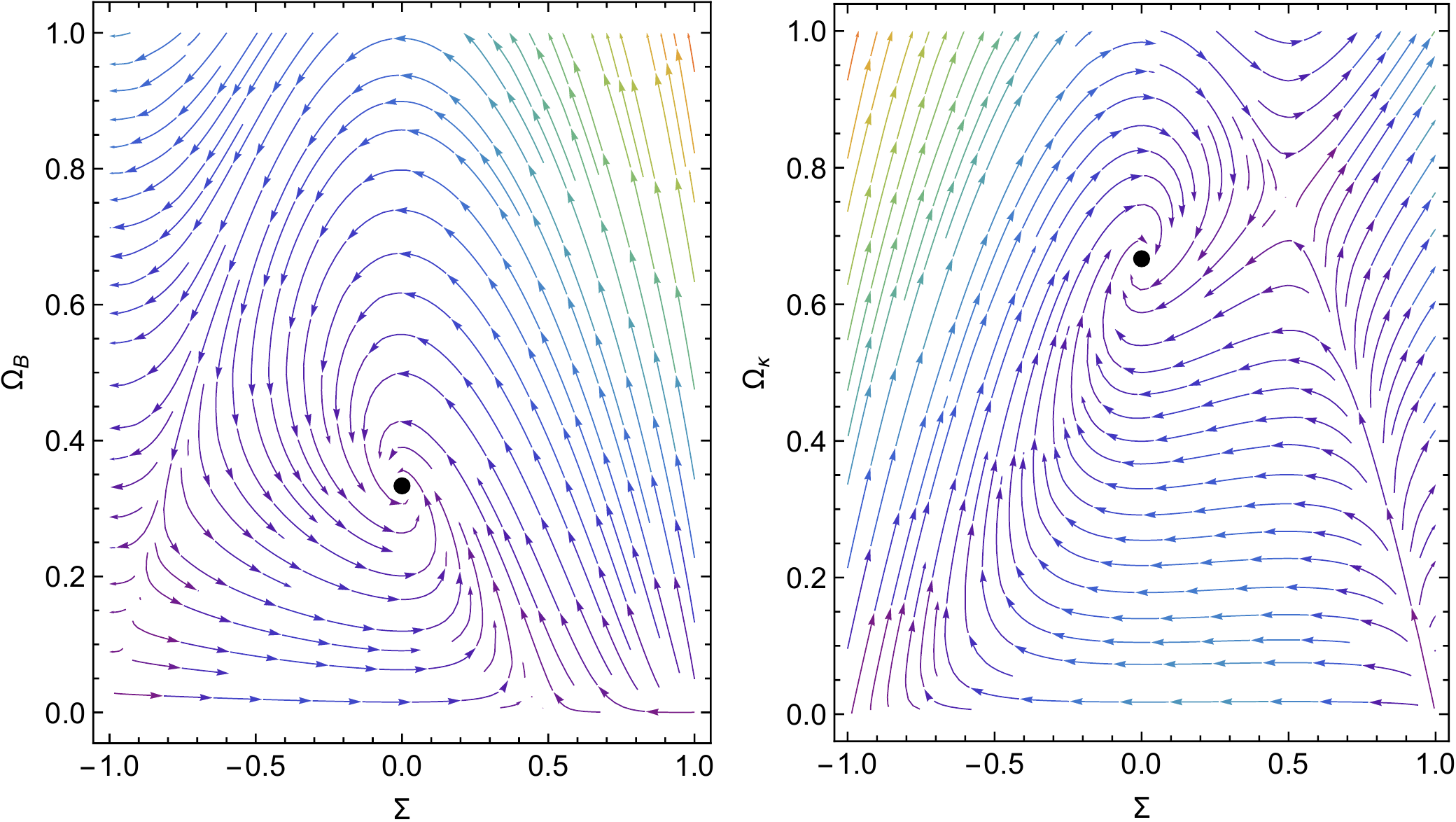}}
\vspace*{8pt}
\caption{Attractor behavior of the shear-free solution on the planes
$\Omega_B\times\Sigma$ (left) and $\Omega_\kappa\times\Sigma$ (right).\protect\label{fig1}}
\end{figure}

The above system is quite general and can be applied to different scenarios with different perfect 
fluids. We are particularly interested to see whether an inflationary (more precisely, de Sitter) 
phase would produce shear-free expansion, which would then determine the metric during 
the following radiation and matter dominated eras, possibly affecting the formation of CMB 
anisotropies. We thus consider the case with $\omega=-1$, for which the point
\be
(\Sigma,\Omega_B,\Omega_\kappa) = (0,\,1/3,\,2/3)\,,
\ee
is a stable fixed point~\cite{MPPT:2016ab} -- see Fig.~\eqref{fig1}. It is worth 
mentioning that, since by definition $\Omega_\kappa\propto -\kappa$ (see~\eqref{omegas}), 
the above result imply that only the BIII geometry is dynamically stable. However, this does not 
exclude the possibility that the KS geometry leads to a stable fixed points when couplings between 
the fluids are allowed. For a more sophisticated dynamical analysis in KS spacetimes, 
see~\cite{Fadragas:2013ina}.

Thus, cosmologies with metric~\eqref{b3ks} possess an attractor solution in which the universe, 
although anisotropic, will expand isotropically. Since the metric has only one scale factor, it can 
be brought to a conformally static form
\be
\dd s^2 = 
a^2(\eta)\left[-\dd\eta^2+\gamma_{ab}(x^c)\dd x^a\dd x^b+\dd z^2\right]\,,
\label{shear-free-metrics}
\ee
which implies that the CMB will be perfectly isotropic~\cite{Clarkson:1999yj,Ehlers:1966ad}. 
Moreover, provided that the stress tensor ``balances'' the spatial curvature\footnote{Rigorously 
speaking, the stress tensor has to equal the electric part of the Weyl tensor in these 
spacetimes~\cite{Mimoso:1993ym}.} -- see Eq.~\eqref{balance} -- the shear decays and the 
background equations in conformal time will be given by
\begin{align}
\mathcal{H}^2 & = \frac{1}{3}a^2\rho - \frac{\kappa}{a^2}\,,\\
\mathcal{H}' & = -\frac{1}{6}a^2(\rho + 3p)\,.
\end{align}
These are exactly the Friedmann-Robertson-Walker (FRW) equations of universes with spatial 
curvature. Evidently, the anisotropy in the spatial curvature will lead to new signatures at the 
perturbative level, which we now explore.

\subsection{Perturbation Theory}\label{perturbation}

As far as the machinery of gauge-invariant and linear cosmological perturbations is concerned, 
perturbation theory in anisotropic 
spacetimes~\cite{Pereira:2007yy,Gumrukcuoglu:2007bx,Tomita:1985me,Den:1986ac} is essentially the 
same as its isotropic cousin~\cite{Mukhanov:1990me}. Nonetheless, when one departures 
from FRW universes, there are three main aspects which require attention. These are: 
\begin{enumerate}
 \item the dynamics of the background spacetime;
 \item the geometry of constant-time hypersurfaces;
 \item the determination of spatial eigenfunctions.
\end{enumerate}
Thus, for example, perturbation theory in Bianchi I spacetimes is directly affected by item (1) 
since, at the background level, the anisotropy of expansion couples perturbative modes through the 
background shear, even if they are decoupled at some initial 
time~\cite{Pereira:2007yy,Gumrukcuoglu:2007bx}. Consequently, one cannot track the evolution of 
each perturbative mode independently, which considerably complicates the analysis. SF models are 
obviously exempt from this difficulty, and in this regard they are much simpler to perturb. On the 
other hand, SF are directly affected by items (2) and (3), from which most of their distinctive 
observational signatures follows.

SF space-times are orthogonal models, which means that they admit a timelike vector field 
everywhere orthogonal to the spatial hypersurfaces. Thus, say, metric perturbations can be 
naturally split into time-time (or scalars, e.g. $\delta g_{00}$), 
space-time (or 3-vectors, e.g., $\delta g_{0i}$), and space-space components (or 3-tensors, e.g. 
$\delta g_{ij}$). Next, 3-vectors and 3-tensors can be further decomposed into their irreducible 
pieces, according to the symmetries of the spatial hypersurfaces where they live. In the case of 
isotropic FRW spacetimes, the spatial sections are invariant under the SO(3) group, which leads to 
the standard Scalar-Vector-Tensor (SVT) decomposition of 
perturbations~\cite{Mukhanov:1990me,Bertschinger:1993xt}. As we have seen, the spatial hypersurfaces 
of SF universes is a product manifold, and we thus implement an irreducible decomposition 
according to the symmetries of each submanifold. 
%%%%%%%%%%%%%%%%%%%%%%%%%%%%%%%%%%%%%
\begin{table}[h]
\tbl{Comparison of mode-splitting in different space-times.}
{\begin{tabular}{@{}cccc@{}} \toprule
Space-time & Spacetime splitting & Irreducible pieces \\
\colrule
FRW\hphantom{00} & \hphantom{0}1+(3) & \hphantom{0}Scalar + Vector + Tensor \\
BIII/KS\hphantom{00} & \hphantom{0}1+(2+1) & \hphantom{0}Scalar + Vector + Scalar \\
\colrule
\end{tabular}\label{ta1} }
\end{table}
%%%%%%%%%%%%%%%%%%%%%%%%%%%%%%%%%%%%%

Thus, in the real line $\mathbb{R}$ one can only have scalars, while in the two-dimensional 
submanifold of metric~\eqref{shear-free-metrics}, vectors and tensors are 
decomposed as:
\begin{align}
V^a & = D^aV + \bar{V}^a\,,\\ 
h^{ab} & = 2S\gamma^{ab}+D^aD^bU + D^{(a}\bar{E}^{b)}\,,
\end{align}
where $(V,S,U)$ are scalars and $(\bar{V}^a,\bar{E}^a)$ are transverse vectors: 
$D^a\bar{V}_a=0=D^a\bar{E}_a$. Note that, in two dimensions, transverse vectors are essentially 
scalars. Moreover, there are no transverse and traceless tensors in two dimensions. Of course, this 
does not mean that there are no gravitational waves, but rather that each polarization of the wave 
come from a different perturbative sector. However, this does imply that each polarization will 
have its own dynamics. Incidentally, this is a general feature of anisotropic 
spacetimes~\cite{Pitrou:2008gk,Pereira:2012ma} which can be relevant to the recently founded era of 
gravitational wave astronomy~\cite{Abbott:2016blz}.

The last step to the implementation of perturbation theory is the determination of a complete set 
of spatial eigenfunctions. This step cannot be overlooked since, without it, crucial cosmological 
observables, like the primordial power spectrum, cannot be computed. The eigenfunctions 
$\phi_{\mathbf{q}}$ that we are looking for are the solutions of the eigenvalue problem
\be
\frac{1}{\sqrt{h}}\partial_i\left(\sqrt{h}\partial^i\phi_{\mathbf{q}}\right)=-q^2\phi_{\mathbf{q}}\,
,
\ee
where $h_{ij}$ is the metric on spatial sections of~\eqref{shear-free-metrics}. In the cylindrical 
coordinates of metric~\eqref{b3ks}, these eigenfunctions can be found by means of a simple 
separation of variables. They are, up to a normalization factor, given 
by~\cite{Pereira:2015pxa,Adamek:2010sg}:
\be
\phi_{\mathbf{q}}(\mathbf{x})\propto\begin{cases}\label{eigenf}
P_{-1/2+i\ell}^{m}(\cosh\rho)e^{im\varphi}e^{ikz}\,, & 
\qquad(\rm{BIII})\\
P_{\ell}^{m}(\cos\rho)e^{im\varphi}e^{ikz}\,, & \qquad(\rm{KS})
\end{cases}
\ee
where $P_{\nu}^\mu(z)$ are associate Legendre polynomials\footnote{Note that 
$(\rho,\ell)\in\mathbb{R}^+$ in $\mathbb{H}^2$, whereas $\rho\in[0,\pi)$ and 
$\ell\in\mathbb{N}$ in 
$\mathbb{S}^2$.}. The eigenvalues $\ell$, $m$ and $k$ are related to the wave-vector $\mathbf{q}$ 
through the following dispersion relations:
\be\label{dispersion}
q^2=\begin{cases}
\ell^2+k^2+1/4\,, & \qquad(\rm{BIII})\\
(\ell+1/2)^2+k^2-1/4\,. & \qquad(\rm{KS})
\end{cases}
\ee

Some general remarks about these results are in order: first, it is straightforward to show that, 
in the limit of small distances and large $\ell$, both eigenfunctions become
\[
\phi_{\mathbf{q}}(\mathbf{x})\propto \ell^{1/2}J_m(\ell\rho)e^{im\varphi}e^{ikz}\,.
\]
As expected, these are the spatial eigenfunctions of the Laplacian on a flat FRW universe. Second, 
note that the eigenvalues $m$ do not appear explicitly in the dispersion 
relations~\eqref{dispersion}, which reflects the residual rotational symmetry 
of~\eqref{shear-free-metrics}. Finally, we note 
that in both BIII and KS cases there is an intrinsic lower limit to the ``Fourier'' mode $q$. In 
fact, the largest wave in BIII has $\ell=0=k$, whereas in KS it has\footnote{The case $\ell=0$ in 
KS corresponds to a monopole, and can thus be neglected.} $\ell-1=0=k$. In both cases, thus
\be
q\geq\frac{1}{|{\rm curvature\;\;scale}|}\,.
\ee
In other words, there can't be a wave larger than the curvature scale in such universes. This 
feature offers an interesting observational window through the Grishchuk-Zel'dovich 
effect~\cite{grishchuk1978long,GarciaBellido:1995wz}.

From the above recipes, it is a straightforward but rather tedious task to parameterize metric and 
matter perturbations, construct gauge-invariant variables and linearize Einstein equations. The 
reader interested in the details can check Ref.~\refcite{Pereira:2012ma}. We now comment on the 
observational signatures that SF models would imprint in the CMB.

\subsection{Observational signatures}\label{signatures}
In order to discuss observational signatures of SF models, let us thus focus on scalar 
perturbations. Assuming an inflationary period, the perturbations to the metric are found to 
be~\cite{Pereira:2015pxa}:
\be
\dd s^2 = a^2(\eta)[-(1+2\Phi)\dd\eta^2 + (1-2\Phi)\dd x_i\dd x^i]\,,
\ee
where $\dd x_i\dd x^i=\gamma_{ab}\dd x^a\dd x^b + \dd z^2$ and where $\Phi$ is the only 
gauge-invariant scalar metric perturbation. Interestingly, the above line element is identical to 
the one we find from scalar metric perturbations in an inflationary FRW universe. Thus, from the 
dynamical point of view, $\Phi$ has the same time evolution as the Newtonian gravitational of 
standard perturbation theory. A corollary of the above result is that, since the Sachs-Wolfe (SW) 
effect is purely kinematic, it has the same shape in SF universes. That is
\be
\label{sw}
\Delta T(\hat{\mathbf{n}})= \frac{1}{3}\Phi(\mathbf{x},\eta)\,.
\ee
Of course, SW effect will still lead to different signatures, since in the above relation 
$\mathbf{x}$ are the coordinates of a point in a manifold with anisotropic spatial curvature. In 
order see how these differences comes about, we can compute the two-point correlation function 
(2pcf), 
$C(\hat{\mathbf{n}},\hat{\mathbf{n}}')=\langle\Delta T(\hat{\mathbf{n}})\Delta 
T(\hat{\mathbf{n}}')\rangle$, under the assumption that the gravitational potential that we measure 
today is one realization of a Gaussian random variable. In ``Fourier'' space, this is implemented by 
the relation
\be
\langle\Phi(\mathbf{q})\Phi^*(\mathbf{q}')\rangle = 
{\cal P}(\mathbf{q})\times\begin{cases}
(\tanh{\ell\pi})^{-1}\delta_{mm'}\delta(\ell-\ell')\delta(k-k') & \qquad(\mbox{BIII})\,,\\
(2\pi)\;\delta_{mm'}\delta_{\ell\ell'}\delta(k-k') & \qquad(\mbox{KS})\,,
\end{cases}
\ee
where ${\cal P}(\mathbf{q})$ is the (anisotropic) primordial power spectrum. Expanding $\Phi$ in 
the eigenfunctions~\eqref{eigenf} and using some identities of Bessel functions, one can easily 
show that, in the BIII case~\cite{Pereira:2015pxa}
\be\label{2pcfSF}
C(\hat{\mathbf{n}},\hat{\mathbf{n}}') = 
\frac{1}{(6\pi)^2}\begin{cases}
\int_0^\infty\ell\dd\ell 
\int_{-\infty}^{+\infty}\dd k\,{\cal P}(\ell,k)P_{1/2+i\ell}(\cosh\Delta\rho)e^{ik\Delta 
z}\,,\quad\mbox{(BIII)}\\
\sum_{\ell=1}^\infty\left(\ell+\frac{1}{2}\right) 
\int_{-\infty}^{+\infty}\dd k\,{\cal P}(\ell,k)P_{\ell}(\cos\Delta\rho)e^{ik\Delta 
z}\,,\quad\mbox{(KS)}
\end{cases}
\ee
where 
$\cos({\rm h})\Delta\rho\equiv\cos({\rm 
h})\rho\cos({\rm h})\rho'\mp\sin({\rm h})\rho\sin({\rm h})\rho'\cos(\varphi-\varphi')$, with the 
minus\textbackslash plus sign corresponding to the $\cosh\Delta\rho$\textbackslash$\cos\Delta\rho$ 
cases, respectively. We 
remind the reader that, in deriving these expressions, we have used the fact the ${\cal 
P}(\mathbf{q})={\cal P}(\ell,k)$ cannot depend on the eigenvalue $m$, since the latter is associated 
with an angular variable of a rotationally symmetric (sub) space. It is important to compare the 
above two-point functions with the same quantity in FRW universes. In cylindrical coordinates, the 
latter is~\cite{Pereira:2015pxa}
\be
\label{2pcfFRW}
C(\hat{\mathbf{n}},\hat{\mathbf{n}}') = \frac{1}{(6\pi)^2}\int_0^\infty\ell\dd\ell 
\int_{-\infty}^{+\infty}\dd k\,{\cal P}(q)J_{0}(\ell\Delta\rho)e^{ik\Delta z}\,,
\ee
where $\Delta\rho^2=\rho^2+\rho'^2-2\rho\rho'\cos(\varphi-\varphi')$. A further 
integration\footnote{If we let $\ell=q\sin\psi$ and $k=q\cos\psi$, the integral in $\psi$ can be 
evaluated analytically, leading to the famous expression for the 2pcf in real-space. See 
Ref.~\refcite{peebles1980large}.} reveals that, in this case, the 2pcf is only a function of 
$\vartheta=\arccos(\mathbf{\hat{n}}\cdot\mathbf{\hat{n}}')$, as one expects from isotropy. However, 
we prefer to keep Eq.~\eqref{2pcfFRW} in its present form to compare with 
Eqs.~\eqref{2pcfSF}. There are two main differences between them. The first is 
obviously in the anisotropy of the primordial power spectrum ${\cal P}$, which in the SF cases is a 
function of the modes $\ell$ and $k$. Rigorously speaking, one should fix ${\cal P}(\ell,k)$ by 
canonically quantizing the inflaton perturbations in SF universes. However, since we are only 
interested in the general signatures of SF models, we can take a simpler route by demanding that 
the anisotropic 2pcf recovers $C(\vartheta)$ in the limit of coincident points:
\be
\left.C(\mathbf{n},\mathbf{n}')\right|_{\mathbf{n}=\mathbf{n}'}\overset{!}{=}C(\vartheta)
\ee
which completely fix ${\cal P}(\ell,k)$ in terms of the isotropic power spectrum ${\cal P}(q)$. 
The second difference between~\eqref{2pcfFRW} and~\eqref{2pcfSF} is that the function $J_{0}$ in the 
kernel of the integral gets replaced by $P^\mu_{\nu}$, the latter being a function of the distance 
between two points in a curved two-dimensional space. As we know, CMB data suggest that the 
observable universe is spatially flat~\cite{Ade:2015xua}. Thus, in the light of current 
observations, shear-free models should be considered in the limit of large spatial curvature, where 
$P^{\mu}_\nu\rightarrow J_0$. Evidently, we are interested in the next-to-leading order corrections, 
since those will lead to statistical anisotropies in CMB maps. Schematically, thus, we can 
write the large-curvature limit of~\eqref{2pcfSF} as~\cite{Pereira:2015pxa}:
\be
\label{large-curvature}
C(\hat{\mathbf{n}},\hat{\mathbf{n}}')=C(\vartheta)\pm{\cal F}(\hat{\mathbf{n}},\hat{\mathbf{n}}')
\ee
where the plus\textbackslash minus signs correspond to BIII\textbackslash KS cases, respectively, 
and the function ${\cal F}$ is of second order in the quantity $(\Delta\eta/L)$, that is, the 
horizon distance $\Delta\eta$ in 
units of the curvature scale $L$. Since the latter is not known, this ratio appears as a free 
parameter in the model. 
%%%%%%%%%%%%%%%%%%%%%%%%%%%%%%%%%%%%%%%%%%%%%
\begin{figure}[th]
\centerline{\includegraphics[scale=0.45]{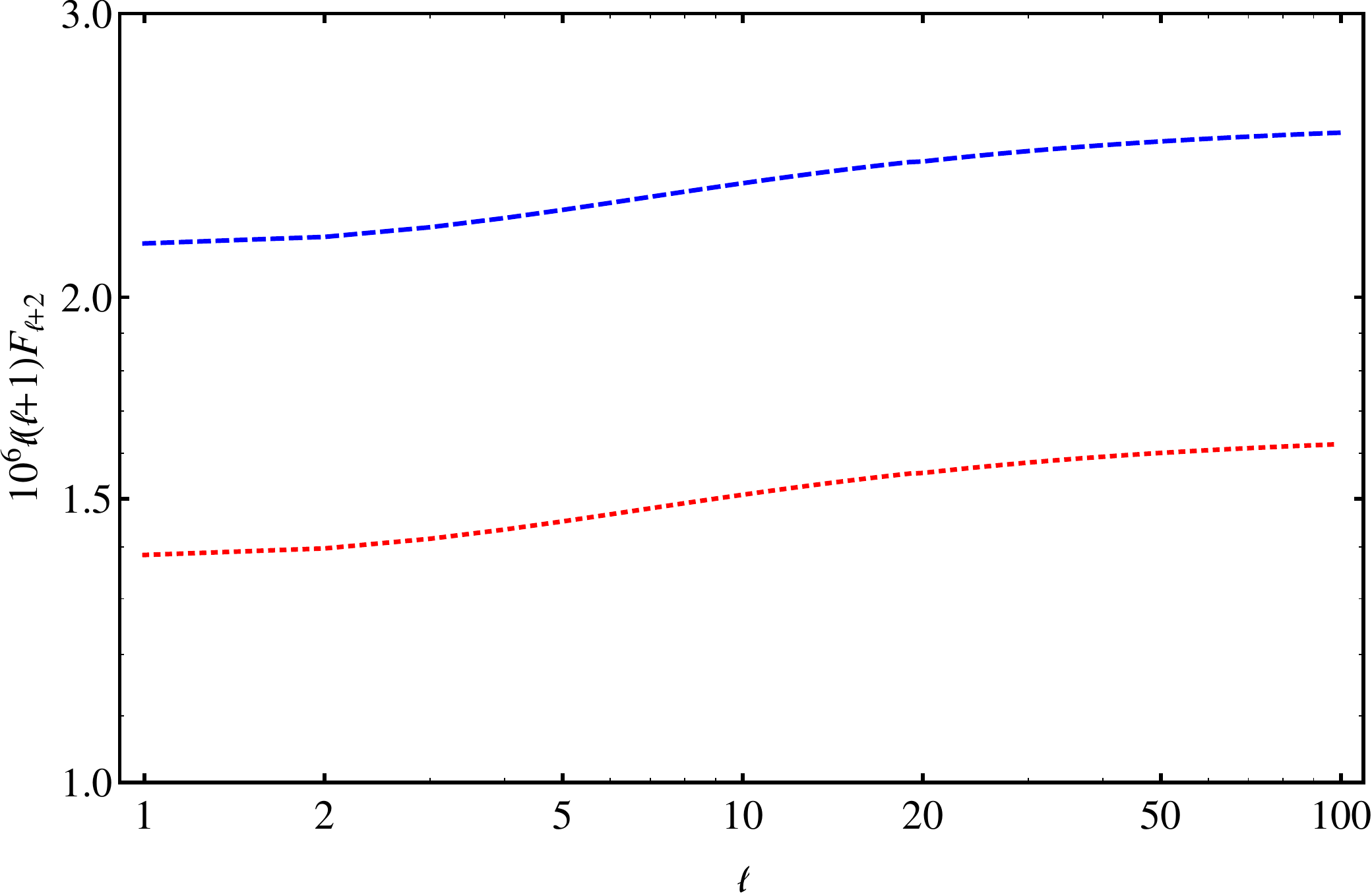}}
\vspace*{8pt}
\caption{Di-log plot of the quantity $\ell(\ell+1){\cal F}_{\ell+2}$\protect\label{fig2}. These 
curves correspond to $\Delta\eta/L=0.04$ (red, dotted curve) and $\Delta\eta/L=0.04$ (blue, dashed 
curve).}
\end{figure}
%%%%%%%%%%%%%%%%%%%%%%%%%%%%%%%%%%%%%%%%%%%%%
One interesting aspect of the function ${\cal F}$ is that it only couples multipoles of 
equal parities, which is a direct consequence of parity invariance of metric~\eqref{b3ks}. In 
order to have an idea of the shape of this function, we define its angular power spectrum in 
terms of off-diagonal terms of the CMB covariance matrix:
\be
{\cal F}_{\ell+\Delta\ell} \equiv \frac{1}{2\ell+1}\sum_{m=-\ell}^\ell|\langle a^*_{\ell 
m}a_{\ell+\Delta\ell,m}\rangle|\,,
\ee
which can in turn be directly computed from~\eqref{large-curvature}. The strongest signal is 
expected to come from the closest neighbors in multipole space. Since multipoles with separation 
$\Delta\ell=1$ give zero signal due to parity, the next effect results from neighbors with 
$\Delta\ell=2$. We show in Figure~\eqref{fig2} a plot of this quantity for two arbitrary choices of 
the parameter $\Delta\eta/L$. It is interesting to note that this function grows smoothly with 
growing $\ell$ -- a feature which, in observational terms, might alleviate the cosmic variance of 
very low multipoles.

\section{Conclusions and Perspectives}\label{perspectives}

Unprecedented progress in observational cosmology compel us to explore theoretical 
possibilities beyond the simple scenario of an isotropic and spatially flat universe. However, since 
the framework of a vanilla $\Lambda$CDM cosmology becomes stronger after each new observational 
mission, it is important to develop models with new degrees of freedom that do not spoil known 
observational results.

In this work we have explored cosmological models which, starting from a general anisotropic 
configuration, are rapidly attracted to an FRW-like model, while still being anisotropic at
the level of its spatial curvature. Although we have focused on the choice of a 
specific anisotropic source of matter to obtain this feature, the simplicity of our model suggest 
that it might hold in more general scenarios. Since the background dynamics of these 
models is exactly the one of a curved FRW universe, it represents an interesting 
counter example to our (unjustified) intuition that the isotropy of CMB requires an equally 
isotropic universe.

By developing a proper decomposition of perturbative modes in spaces with anisotropic curvature,  
we have shown that linear perturbation theory in shear-free models is perfectly doable, and leads 
to interesting phenomenological consequences -- one of which is the existence of a geometrical 
upper bound to the wavelength of cosmological perturbations. Assuming that the curvature of the 
universe lurks just beyond the current horizon radius, we have computed off-diagonal signatures that 
an anisotropically curved geometry would imprint on the temperature spectrum of CMB. Such effects 
could be responsible to some of the known CMB statistical anomalies, although further investigation 
is required to clarify this issue. 

Finally, we comment on an interesting possibility to extend the results of this work. A general 
prediction of anisotropic cosmological metrics, and of shear-free metrics in particular, is 
that each polarization of gravitational waves should have its own dynamics. This will be an 
interesting signature to look for in the forthcoming measurements of primordial gravitational 
waves\footnote{http://www.core-mission.org/}. Moreover, due to the specificities of the 
eigenfunctions and mode decomposition in spaces with anisotropic curvature, we also expect the 
tensor-to-scalar ratio to be very different in these models~\cite{FT:2016bc}. Thus, future 
measurements of the $B$-modes in the CMB polarization maps will offer a new window to constrain both 
the isotropy and the curvature of the universe.

\section*{Acknowledgments}

We would like to thank Martin Bucher and Yeinzon Rodriguez for their kind invitation to write this 
review, and also for being flexible with the deadlines. We also thank Yeinzon Rodriguez, Juan Pablo
Beltr\'an, Cesar Alonso Valenzuela and Juan Carlos S\'anchez for the superb workshop they organized 
at Cali. This work was partially supported by Conselho Nacional de Desenvolvimento Científico e 
Tecnológico (CNPq) under grants 485577/2013-5 and 311732/2015-1.

%\section*{References}

%They are to be cited in the text in superscript
%after the punctuation marks e.g.~word,\cite{Marnelius} and word:\cite{Bjorken}
%If it is mentioned in the text as part of a sentence, it should be of normal size,
%e.g.~see Ref.~\refcite{Bohr}. Please list using the style shown in the following examples.
%For journal names, use the standard abbreviations. Typeset references in 9 pt Times Roman.
%Each reference number should consist of one reference only.

\bibliographystyle{ws-mpla}
\bibliography{shear-free-refs}

\end{document}